\documentclass[]{rstransa}
\usepackage{multicol}
\usepackage{nicefrac}

\newcommand{\new}[1]{{#1}}

\DeclareMathOperator{\isi}{ISI}
\DeclareMathOperator{\var}{Var}

\begin{document}

\title{The broad edge of synchronisation: Griffiths effects and collective phenomena in brain networks }

\author{
Victor Buend{\' i}a $^{1,2}$, Pablo Villegas $^3$, Raffaella Burioni $^{4,5}$ and Miguel A. Mu\~noz$^{6}$}

\address{
$^{1}$ Max Planck Institute for Biological Cybernetics, T\"ubingen, Germany.
$^{2}$ Department of Computer Science, University of T\"ubingen, T\"ubingen, Germany.
$^{3}$ IMT Institute for Advanced Studies, Piazza San Ponziano 6, 55100 Lucca, Italy.
$^{4}$
Dipartimento di Matematica, Fisica e Informatica,  Universit\`a di Parma, via G.P. Usberti, 7/A - 43124, Parma, Italy.
$^{5}$ INFN, Gruppo Collegato di Parma, via G.P. Usberti, 7/A -
  43124, Parma, Italy.
$^{6}$
Departamento de Electromagnetismo y F{\'i}sica de la Materia
  e Instituto Carlos I de F{\'i}sica Te{\'o}rica y
  Computacional. Universidad de Granada, E-18071 Granada, Spain.
}
\subject{Physics, Network theory, Neuroscience}


\corres{Miguel A. Mu{\~n}oz\\
\email{mamunoz@onsager.ugr.es}}

\begin{abstract}
  Many of the amazing functional capabilities of the brain are collective properties stemming from the interactions of large sets of individual neurons. In particular, the most salient collective phenomena in brain activity are oscillations, which require the synchronous activation of many neurons.  Here, we analyse parsimonious dynamical models of neural synchronisation running on top of synthetic networks that capture essential aspects of the actual brain anatomical connectivity such as a hierarchical-modular and core-periphery structure. These models reveal the emergence of complex collective states with intermediate and flexible levels of synchronisation, halfway in the synchronous-asynchronous spectrum. These states are best described as broad Griffiths-like phases, i.e. an extension of standard critical points that emerge in structurally heterogeneous systems. We analyse different routes (bifurcations) to synchronisation and stress the relevance of ``hybrid-type transitions'' to generate rich dynamical patterns.  Overall, our results illustrate the complex interplay between structure and dynamics, underlining key aspects leading to rich collective states needed to sustain brain functionality.
 \end{abstract}






\maketitle

\section{Collective effects in spontaneous brain activity}

The human brain is the most complex object we are aware of.  It is composed of as many as $10^{11}$ neurons and $10^{15}$ synapses forming an extraordinarily intricate network \cite{Kandel,Sporns-book,Fornito}.  Electrochemical signals traveling \new{through such a network} permit neurons to communicate with each other, allowing them to coordinate their behavior and generate a vast diversity of possible emergent or collective behaviours, which are believed to be at the basis of information transmission, processing, and storage \cite{Dayan}.  Even if individual neurons such as the \emph{Jennifer-Aniston's} one ---responding with high specificity to images of the famous actress--- exist \cite{Jennifer}, these are more the exception than the rule: most of the remarkable abilities of the brain are delocalised, i.e. they emerge out of \emph{collective phenomena} stemming from the interactions of large sets of neurons \cite{Bialek-book,Bialek-collective,RMP,Taglia-collective}.

Probably, the most salient collective phenomena in brain activity are oscillations, as already detected in pioneering electroencephalogram recordings almost one century ago \cite{Buzsaki}. Neural oscillations at a given (mesoscopic) brain region require of the synchronous firing or activation of many individual neurons within it, allowing signals to robustly propagate to e.g. distant network areas \cite{Buzsaki,Battaglia-synchro}. Indeed, neural synchronisation has been shown to be crucial in memory, attention, vision, and high-level cognitive functions, and abnormalities in the level of synchrony have been associated with pathologies such as, e.g., Parkinsonian disease (excess) and autism (deficit) \cite{Parkinson,autism}. In what follows, we briefly  review  four relevant concepts useful to shed light onto whole-brain connectivity and dynamics.

{\bf A) Resting-state functional networks.} Brain networks are never silent; even in resting conditions their ongoing activity consumes as much as $20\%$ of our total energy budget, suggesting crucial functional roles \cite{Raichle}.  Seminal work developed in the last two decades guided by novel neuroimaging technologies elucidated that \new{spontaneous activity  in resting conditions at different locations} (as measured, e.g., by functional magnetic resonance imaging (fMRI) techniques) does not occur in a random uncorrelated way.  Even if the resulting activity time series at diverse mesoscopic regions are non-stationary (see below), most of the first studies focused on the analysis of long-time averaged pairwise correlations. In this way, regions of correlated and anticorrelated activity were robustly identified, defining the so-called ``resting-state networks'' (RSN) \cite{RSN1,RSN2,RSN3,RSN4,Fox,biswal2012}, which are believed to represent the basic functional architecture of the brain \cite{Sporns-book,Fornito,Bassett-review, Review-van}.  These networks of functional correlations show a rich multi-scale organisation with moduli of correlated activity appearing within other moduli at larger scales in a nested hierarchical-modular way \cite{Meunier,Zhou-hierarchical,Ibai,Gabrielli-hierarchical}.

The focus of research in resting-state functional connectivity has recently shifted from {\emph static} (i.e. long-time-averaged correlations) to {\emph dynamical} aspects (in the sense that correlations are measured independently in a series of concatenated time windows), thus allowing for a characterisation of the temporal organisation of functional connectivity or \emph{chronnectomics} \cite{Glover,Calhoun,Dynamic-FC,Breakspear2014}. The emerging picture is that \new{the global patterns of spatial correlations} shift in time between a discrete set of basic functional (metastable) states \cite{Deco-discrete1,Deco-discrete2}, thus creating a rich dynamical repertoire which is believed to constitute a template to sustain diverse functions. 

{\bf B) Structural networks.}  Resting-state functional correlations are well-understood by now to be strongly constrained by the underlying network of structural (or anatomical) connectivity (as determined from diffusion tensor imaging (DTI) measurements combined with algorithms for the identification of white-matter fiber tracts connecting the units of a given mesoscopic parcellation \cite{Fornito}).
Important architectural features of the resulting structural networks (or \emph{structural connectomes}) at the mesoscale are that: (i) they are organised in a hierarchical-modular way \cite{Sporns2014,Review-Bullmore,Sporns-book,Fornito,SCKH} and, (ii) they have a core-periphery organisation with connector hubs \cite{Eguiluz,Bassett-core,Bassett-core2}. Actually, areas with a strong mutual structural connectivity show also a high level of long-time functional connectivity \cite{Hagmann,Honey09}. Reciprocally, functionally connected regions are likely to be directly wired together \cite{van-2009,Ibai}, revealing that structural networks constitute a scaffold to support and constrain the emergence of functional correlations. However, there is not a one-to-one equivalence between structure and function and, \new{as mentioned above,} the focus of recent research is on time-dependent aspects of functional connectivity. 

{\bf C) Segregation-and-integration balance}
An influential idea, proposed by Tononi and coworkers, is that high-level cognitive tasks performed by the brain require \emph{segregated} processing for each type of (e.g., sensory) input, which then needs to be \emph{integrated} to allow for a unified representation, advanced cognitive processing, and response \cite{Tononi}. Thus, it was conjectured that an optimal balance between segregation and integration is crucial for optimal brain function and that deviations from such a balance are associated with neurological disorders \cite{Tononi}. Furthermore, such an optimal and flexible balance requires high diversity and variability of underlying synchronisation or correlation patterns of neural activity to be sustained \cite{Dynamic-brain, Carhart-entropic}. Thus, it is important to shed further light on what are the (i) structural and (ii) dynamical aspects favouring an optimal segregation-integration balance, i.e. flexible and variable levels of synchronisation.

(i) \emph{From the structural side}, it is well-established that a hierarchical-modular network organisation is particularly well suited to accommodate diverse levels of segregated/integrated activity across many scales: local moduli support segregation but the transient   correlation or synchronisation of  some of them allows for integration of information at progressively larger scales in a hierarchical way \cite{Sporns-SI,Tononi-SI,Deco-SI,Zhou-jointly}. Similarly, the presence of central connector hubs and the resulting core-periphery structure has been reported to be crucial for the emergence of a central integrative core controlling the segregation-integration balance \cite{broken-balance,Gollo-core}.

(ii) \emph{From the dynamical side}, ambitious computational models for the ``whole-brain dynamics'' have been constructed in recent years \cite{Cabral-review,Breakspear-review}. These models aim to describe the overall patterns of brain activity in different states (resting,  awake, anesthesized, etc.)
\new{in a parsimonious way}.
They consist on an empirically-determined structural connectome network, \new{where each single node is described by a dynamical unit} (e.g. a neural-field model such as the Wilson-Cowan one \cite{WC,Atasoy1} or a simpler effective model of neural synchronisation such as a Stuart-Landau oscillator \cite{Deco-Hopf,Cabral-review}) assumed to capture key aspects of neural activity at a
 brain mesoscopic region, and \new{ the different units are linked together following the network architecture} \cite{Cabral,Deco-Jirsa,Deco-Hopf,Deco-whole-brain}.
These analyses revealed that \new{the best agreement between model-generated correlations and  empirically-measured ones} is obtained if the dynamics at individual mesoscopic nodes operates close to a (Hopf) bifurcation point \cite{Deco-Hopf,Cabral-review}. This means that high levels of overall dynamical variability, as those reported in the actual resting-state functional networks, are best reproduced if each mesoscopic unit is at the edge of becoming oscillatory.

{\bf D) Criticality and Griffiths phases.} This last result resembles a much-discussed and suggestive conjecture proposing that the large variability of brain activity at vastly different scales might stem from an underlying dynamical process operating at a critical point, i.e., at the edge between order and disorder, with its concomitant spatio-temporal scale-invariance \cite{BP2003,Chialvo2010,Schuster,Breakspear-review,RMP}. The so-called ``criticality hypothesis'' sparked much interest, as critical behavior has been shown to entail many potential functional advantages such as maximal dynamical range, exquisite sensitivity to signals, vast information processing and storage capabilities, etc. \cite{Plenz-functional,RMP}. \footnote{This idea
  is also linked to the ``entropic brain'' hypothesis
  \cite{Carhart-entropic,Atasoy+Carhart}.} Importantly, it has been proposed that the concept of criticality ---which ultimately requires parameter fine-tuning or some self-organisation mechanism to be achieved \cite{SOC+SOB,RMP}--- can be made more generic or robust by invoking \emph{Griffiths phases} \cite{GPCN1,GPCN2,Moretti}. These are extended regimes ---sharing many of the remarkable features of critical points--- that appear in physical systems with some inherent structural heterogeneity (such as actual brain networks) \cite{Moretti,Odor1,Odor2}. For instance, as illustrated in Fig. \ref{Sketch}A, a Griffiths-like phase
emerges in a dynamical model of phase synchronisation running on top of a hierarchical-modular network, \new{i.e. a human connectome network, and it is characterised by a broad set of parameter values}  with intermediate, variable, and transient levels of \emph{``frustrated synchronisation''} \cite{Villegas1,Villegas2,Odor2}. This is in contrast to the standard picture observed for the same dynamical model running on more homogeneous random networks where large variability is encountered only in a small neighbourhood around the critical point (see Fig. \ref{Sketch}A). Thus, our group conjectured that Griffiths-like phases could lie at the basis of the remarkable variability of complex synchronisation patterns in actual brain dynamics, thus facilitating a flexible balance between segregation and integration \cite{Villegas1,Villegas2,Odor2}.

Summing up: even if our knowledge on the nature of both structural and functional brain networks has been remarkably enhanced in recent years, there are still key general questions that remain unanswered. In particular, we ask here: what are the crucial aspects of (i) the  structural architecture of brain networks and, (ii) the dynamics at individual mesoscopic nodes that determine \emph{together} the largest variability of synchronisation patterns, thus allowing for a flexible balance between segregation and integration? What is the role played by different types of criticality, i.e. diverse bifurcation types at the individual mesoscopic units? \new{Are Griffiths-like phases relevant to describe brain activity?}

\section{Modeling the interplay between structure and dynamics}

To shed further light on the previous questions, here we perform an exploratory research in which we analyse diverse types of synthetic structural networks capturing key aspects of brain connectivity combined with parsimonious dynamical models for neural synchronisation \cite{Phase-models,Acebron,Breakspear-Kuramoto} that, notwithstanding their simplicity, can give rise to a rich collective phenomenology, including a large variety of bifurcation types (not only Hopf ones) at the mesoscopic level \cite{PRR}. 

{\bf Structural-network models.}  We consider a sort of \emph{network of networks}, in which each mesoscopic unit is actually represented by a ``\emph{basal network module}'' including a large cluster of densely-connected nodes.  These basal networks are linked together following three different types of unweighted and undirected architectures with a fixed total number of nodes: {\bf{(i)}} random Erd\H{o}s-R\'enyi (ER) networks \cite{Newman-book}, {\bf (ii)} hierarchical-modular random (HM-R) networks without hubs \cite{Moretti,Zamora}, and {\bf (iii)} hierarchical networks with a core-periphery (HM-CP) structure involving central connector hubs (so that the degree distribution is scale free) \cite{Zamora} (see Methods). In particular, HM-R networks are already known to exhibit Griffiths-like collective phases of frustrated synchronisation \cite{Villegas1,Villegas2,Odor2}, \new{while} less heterogeneous ER networks do not exhibit such phases, \new{and} HM-CP networks have not been so far explored from this perspective. Thus, our main  focus here is on analysing whether the core-periphery structure fosters or hinders the emergence of broad Griffiths-like phases.

\begin{figure}[hbtp]
  \centering\includegraphics[width=5.5in]{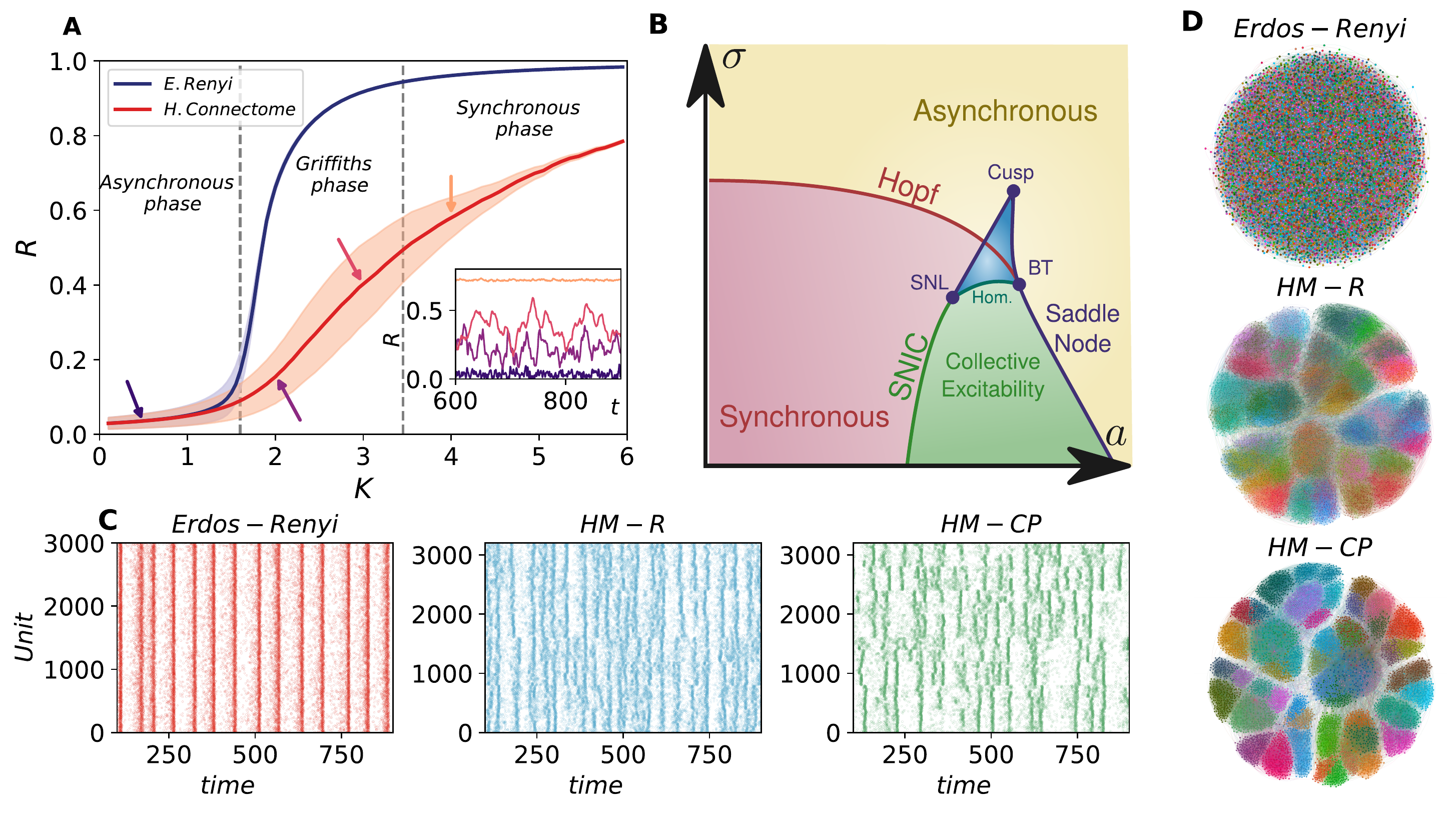}
  \caption{ \textbf{Structural and dynamical ingredients for dynamical richness.}
 {\bf A)} Mean value (continuous line) and standard deviation (shaded region) of the overall synchronisation order parameter, $R$, as measured for the standard Kuramoto model on top of random Erd\H{o}s-R\'{e}nyi (ER) network and on a human-connectome
 structural network \cite{Sporns} (size $N=998$ in both cases \cite{Villegas1}). In the second case, there is a broad Griffiths-like region of ``frustrated synchronisation'', where the level of synchronisation is very variable, as opposed to the standard ER case in which large levels of variability are only observed in the neighbourhood of the critical point (plot adapted from \cite{Villegas1}). {\bf B)} Schematic phase diagram of the model described by Eq.(1) (with a constant frequency, $\omega$) on infinitely-large random networks.
   The transition between the synchronous and
    asynchronous phases can occur either through a Hopf bifurcation (red line), a
    SNIC bifurcation (green line), or ---as a hybrid-type synchronisation---
    through an intermediate region of bistability (blue region) delimited by three
    co-dimension-2 bifurcations: a Bogdanov-Takens, a
    saddle-node-loop, and a cusp bifurcation, as well as by lines of
    homoclinic and saddle-node bifurcations (see \cite{PRR} and
    refs. therein). {\bf C)} Raster plots illustrating crossings of  each network unit 
through a  given phase value ($2\pi$) when  the
    dynamics operate  at the hybrid-type synchronisation transition
    in ER, hierarchical-nodular random network (HM-R), and
    hierarchical-modular core-periphery (HM-CP) networks; much-richer patterns emerge in the latter cases. {\bf D)} Visual representation of the three types of considered networks. Each dot represent a \new{different} node, \new{while} the color code and \new{spatialization} highlight the network hierarchical modularity (HM-R and HM-CP), or its absence (ER).}
  \label{Sketch} \end{figure}

{\bf Dynamical model.} The standard minimal model for collective synchronisation is the one proposed by Kuramoto \cite{Acebron,Strogatz}: a large set of $N$ individual oscillators, characterised by an intrinsic random frequency \new{and} coupled together following  some connectivity pattern (e.g. a fully-connected network or a random network architecture). Typically, as the coupling strength increases, \new{the} collective state experiences a \emph{Hopf bifurcation} representing the transition from an asynchronous to a synchronous/oscillatory state (i.e. the type of bifurcation considered in the above-mentioned  whole-brain computational models
\cite{Deco-Hopf,Cabral-review}). A more general phase model is: \begin{equation}
  \dot \varphi_j = \omega_j + a\sin\varphi_j + 
  J \sum_{i=1} ^N W_{ij} \sin(\varphi_i - \varphi_j) + \sigma \eta_j(t),
  \label{hybrid}
\end{equation}
where each individual unit, $j$, is characterised by its phase $\varphi_j$ and intrinsic (random) frequency $\omega_j$, $W_{ij}$ is the connectivity matrix, $J$ is the overall coupling strength, and $\eta_j(t)$ is a delta-correlated Gaussian white noise \cite{SK}. The term $a \sin\varphi_j$ accounts for the \emph{excitable nature} of individual units (for $a < \omega_j$ the units oscillate, while for $a >\omega_j$ the units remain trapped in an ``excitable state'' which can be perturbed to become transiently oscillatory by noise or external inputs \cite{Ojalvo}).  In the limiting case $a=\sigma=0$, Eq.(\ref{hybrid}) reduces to the standard Kuramoto model.

Eq. (\ref{hybrid}) has been profusely studied in the literature both for both homogeneous frequencies  as well as  for specific frequency distributions \cite{SK,Childs,Pazo,PRR,Martens}.  Remarkably, as sketched in Fig. \ref{Sketch}B, the emerging dynamics is described by a very rich phase diagram, including synchronous and asynchronous phases, as well as different types of bifurcation lines (or critical points), separating them (the diagram in Fig. \ref{Sketch}B has been theoretically obtained for homogeneous frequencies on a fully connected network \cite{PRR}, but a qualitatively identical one arises when considering a heterogeneous frequency distribution \new{with $a$ and/or $\sigma$ fixed to $0$} \cite{SK,Childs,Pazo,PRR,Martens}).

In this very general phase diagram, there exist two main types of collective synchronisation transitions, i.e. two main ways to enter the synchronous phase in Fig. \ref{Sketch}B: Hopf bifurcations (the fingerprint of ``type-II'' synchronisation) and saddle-node-into-invariant-circle (SNIC) bifurcations (the fingerprint of ``type-I'' synchronisation), which are qualitatively \new{very different from each other} and represent the two main types of synchronisation transitions (see e.g. \cite{Ojalvo}). However, as illustrated in Fig. \ref{Sketch}B in between the lines of bifurcations of these two standard clases, there is also a triangular-shaped region of bistability ---with coexisting states of either low or high activity, respectively--- delimited by three codimension-2 bifurcation points (namely, a Bogdanov-Takens, a saddle-node-loop, and a cusp; for details see \cite{PRR,Childs} and refs. therein).  The detailed discussion of this admittedly complicated phase diagram is beyond the scope of the present brief paper, however, let us just stress the only important aspect for our purposes here: our group has recently uncovered that entering the synchronous phase through such a region of bistability ---i.e. through a so-called ``\emph{hybrid-type}'' synchronisation transition--- a much richer phenomenology, including scale-free avalanches (a typical signature of criticality in neural systems \cite{BP2003,RMP}) emerges \cite{PRR}. This result inspired us to explore the effects that such a hybrid-type transition occurring at the basal moduli could have on the overall macroscopic dynamics of complex brain networks. \new{Does it enhance the overall dynamical richness as compared, e.g., the standard Hopf (type-II) route to synchronisation (as usually employed in above-mentioned whole-brain modelling approaches?}

  Thus, in what follows, we computationally analise Eq.(1) on top of the either ER, HM-R, and HM-CP networks of mesoscopic regions as described above, \new{and we choose parameters as to enter the synchronous regime at each node through either a type-I, a type-II, or a hybrid-type transition.}
  As a first observation, for the sake of visual illustration, let us mention that at the ``hybrid-type'' transition, one can generate very rich dynamical patterns, with partial and transient synchronisation between basal moduli, on hierarchical networks (especially on HM-CP ones) which are not observed on simpler ER networks (see Fig. \ref{Sketch}C).  On the other hand, much simpler patterns appear nearby Hopf or SNIC bifurcations (not shown here but qualitatively similar to those for the ER above; see \cite{PRR}).  The results in this example correspond to the case of homogeneous frequencies (as in \cite{PRR}); nevertheless, given that frequency heterogeneity is an important ingredient in brain dynamics, in what follows, we present much more detailed results for the case of a Gaussian bimodal frequency distribution, $g(\omega)=\frac{1}{2} (\mathcal{N}(-\omega_0,\,\Delta^{2})+\mathcal{N}(\omega_0,\Delta^{2}))$ but fixing $a=0$. The reason why we use this model is that its phase diagram ---which coincides qualitatively with that sketched in Fig. \ref{Sketch}B (but as a function of its own free parameters $\Delta$ and $\omega_0$)--- is exactly known \cite{Martens}, and \new{this helps a lot for fixing parameter values to locate the different bifurcation types}.

\begin{figure}[hbtp]
  \centering\includegraphics[width=1.\columnwidth]{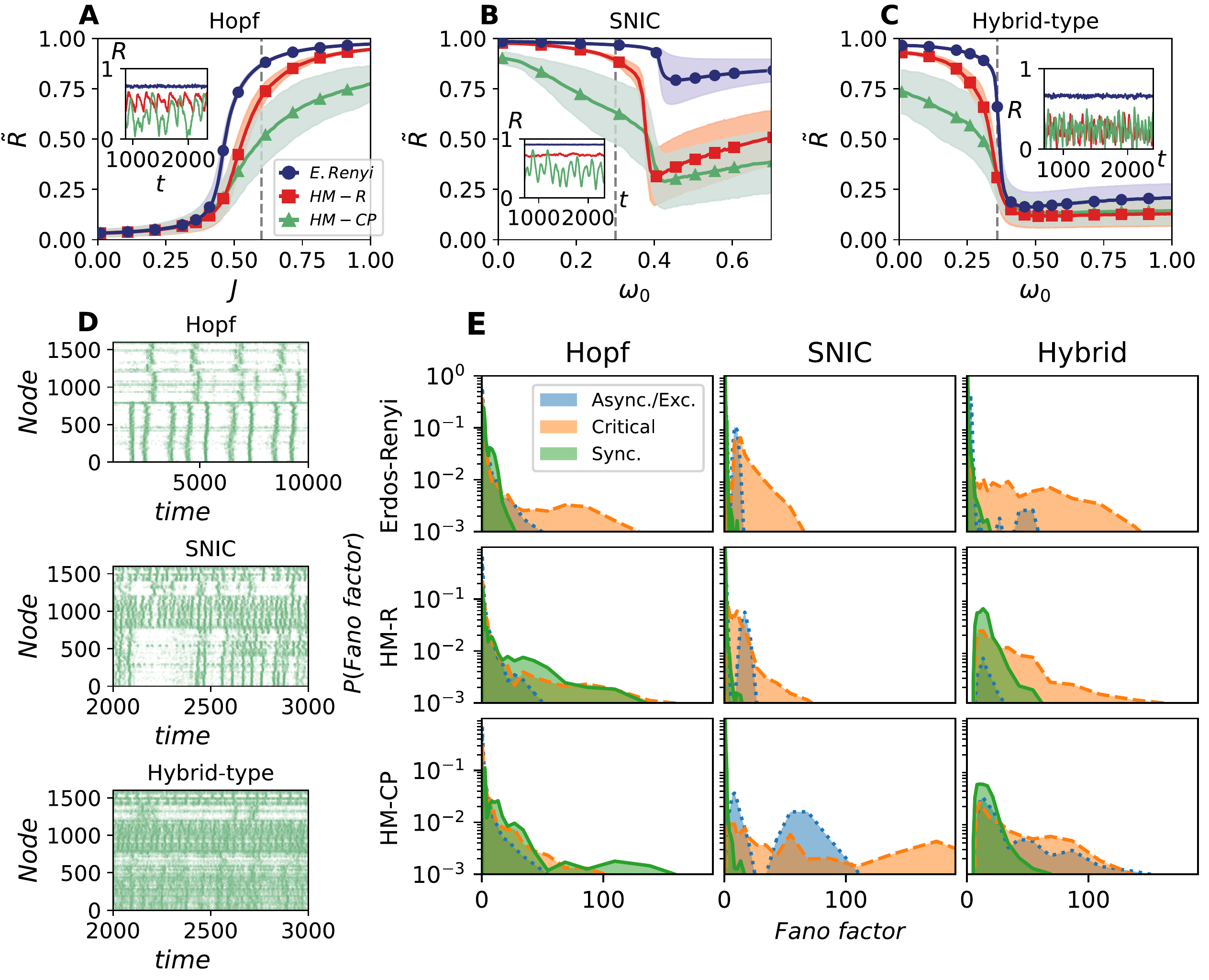}
  \caption{ {\bf Variability across network types and dynamical regimes.}: Mean and standard deviation of the overall synchronisation Kuramoto order parameter, $R$, as measured for the noisy Kuramoto model on top of an ER network (blue circles), a HM-R network (orange squares) and a HM-CP network (green triangles) for the three types of bifurcations: {\bf(A)} Hopf bifurcation \new{($\omega_0=0,\Delta=0.25$), where $J$ acts as control parameter}, {\bf(B)} SNIC bifurcation \new{($J=1,\Delta=0.15$)}, and {\bf(C)} hybrid-type transition \new{($J=1,\Delta=0.28$), where $\omega_0$ acts as control parameter}. The insets show, for every case, some typical temporal series for parameters corresponding to  the vertical dashed line. Observe the emergence of a broad Griffiths-like region of ``frustrated synchronisation'' both for the HM-R and HM-CP networks but not of ER ones. {\bf(D)} Raster plots for the HM-CP network in the three selected bifurcation types: Hopf ($J=0.8$), SNIC ($\omega_0=0.41$) and Hybrid-type ($\omega_0=0.37$). {\bf(E)} Distribution of Fano factors across nodes for different networks and different selected points in the parameters space. (i) Hopf bifurcation: synchronous phase (green solid line, $J=0.6)$, critical point (orange dashed line, $J_{ER}=0.45$, $J_{HM-R}=0.49$, $J_{HM-CP}=0.45$) and asynchronous phase (blue dotted line, $J=0.2$); (ii)  SNIC bifurcation: synchronous phase ($\omega_0=0.3)$, critical point ($\omega_0^{ER}=0.41$, $\omega_0^{HM-R}=0.38$, $\omega_0^{HM-CP}=0.36$) and asynchronous phase ($\omega_0=0.5$);  (iii) Hybrid synchronisation: synchronous phase ($\omega_0=0.3)$, critical point ($\omega_0=0.36$, $\omega_0^{HM-R}=0.35$, $\omega_0^{HM-CP}=0.34$) and asynchronous phase ($\omega_0=0.5$). Other parameter values: $N_{ER}=1500$, $N_{HM-R}=N_{HM-CP}=1600$.
}
\label{networks}
\end{figure}

\section{Results and protocols for further advances}

First of all, to reduce the number of parameter values, we fixed $a=0$, \new{and a small noise amplitude, $\sigma=0.2$, to allow system variability, scrutinising} the system behavior as a function of the remaining free parameters ($\Delta$, $\omega_0$ and $J$). As discussed above, changing one of these two parameters allows one to shift the dynamical regime of basal moduli and encounter different bifurcations types ---Hopf, SNIC or hybrid-type transitions--- separating synchronous from asynchronous regimes (\new{which depends on the ratio $\nicefrac{\Delta}{J}$ and $\nicefrac{\omega_0}{J}$}, see Fig. \ref{Sketch}B and \new{a detailed phase diagram in} \cite{Martens}). First of all, we verified that, very robustly, the synchronous phase becomes a region of ``frustrated synchronisation'' of Griffiths-like phase in both types of hierarchical-modular network (either HM-R and HM-CP)  due to the presence of structural and frequency heterogeneity. In other words, as illustrated in the timeseries showed in the insets of Fig.\ref{networks}A/B/C, the synchronisation Kuramoto order parameter $R$ exhibits high variability (variance) in a broad region in parameter space \cite{Villegas1,Villegas2} (let us remark that such a variability that does not disappear in for infinitely-large networks\cite{Villegas1} ). Note also that similar regimes do not appear in random ER networks (see Fig.\ref{networks}A-C). This observation confirms that hierarchical-modular networks are particularly well-suited to host large dynamical variability and that such a variability is enhanced in networks \new{with the  additional feature of a core-periphery structure}. Observe that also in the ``asynchronous'' phase one can have some residual level of variable synchronisation, especially near the SNIC bifurcation (though this is a spurious effect stemming from the employed bimodal frequency distribution and could be removed by defining a slightly different order parameter).

Let us stress that the Griffiths-like regime is generic across a broad region in phase space and is thus not specific to the type of transition from which such a phase is entered. However, as illustrated in the raster plots of Fig. \ref{networks}D (raster plots are obtained by plotting a dot each time an individual oscillator crosses a predefined value, e.g., $\phi=2\pi$), the qualitative features and variability of the resulting time series within the Griffiths-like phase can vary depending on the bifurcation type near which the system lies.

In order to quantify the levels of variability ---within and across basal moduli--- of the resulting time series we employ the Fano factor ($\text{FF}$), a quantity that takes values $\text{FF} \simeq 0$ for regular ``spikes'', $\text{FF} \simeq 1$ for random Poisson process, and $\text{FF}\gg1$ for highly irregular spikes (see Methods and \cite{Dayan}). Fig. \ref{networks}E shows the probability distribution of Fano factors calculated across individual units, $p(\text{FF})$, for different network types and diverse dynamical regimes, either nearby of far from bifurcation points. For ER networks, large values are observed only if the system is fine-tuned to its critical point while, otherwise, the $\text{FF}$ takes always very low values.  More in general, for all networks the largest values and variability of $\text{FF}$ are typically observed at critical transition points. Nevertheless, observe also that regions of significantly-large variability are obtained within the Griffiths-like phase for HM networks. More specifically, the Fano-factor probability distributions reveal that the largest variability is obtained when the dynamics of basal moduli is tuned to operate near the hybrid-type synchronisation phase transition, with the remarkable exception of the HM-CP: the core-periphery structure always displays a large variability, being maximal near the hybrid transition and the SNIC bifurcation.  \footnote{Previous studies have also employed the Fano factor to assess variability of brain activity (see \cite{Deco-consciousness}); note that the Fano factor can be computed for different observables, and in the cited case it is applied to the variability of instantaneous firing rates over finite-sized temporal windows rather than to inter-spike intervals. Despite of the methodological differences, our results are compatible with the distributions shown in \cite{Deco-consciousness}.}

\new{To shed light on this last observation,} let us remark that the parameters to observe each type of synchronisation transition have been chosen assuming isolated basal moduli. However, external inputs coming from the coupling to other basal moduli can alter the local behavior of each of the basal moduli. Moreover, given the finite size of basal moduli, deviations from the exactly-known mean-field behavior can occur. Also, due to the small size of the hybrid-type transition region (which is not easy to locate numerically \cite{PRR}), combined with finite-network-size effects, could introduce numerical errors in the location of the point of maximal variability.  For all these reasons, even if the hybrid-type transition can in principle ensure the maximal possible variability at a given mesoscopic unit, systems tuned in principle to lie near the SNIC bifurcation seem also to produce very large levels of variability at a global scale (see, e.g., Fig. \ref{networks}D).
 In any case, the previous results confirm the robustness of the Griffiths phase with large degrees of variability for different types of bifurcations (see raster plots and Fano Factor distributions in Fig. \ref{networks}D and E for the HM-CP network).

Thus, in summary, the joint evaluation of phase diagrams and Fano factors support that: (i) The dynamical richness is much more prominent in hierarchical-modular networks than in random ones. (ii) HM-CP networks, including connectors hubs, can sustain more extensive dynamical richness than simple HM-R networks. (iii) In all hierarchical-modular cases, broad regimes of dynamical richness are generated across the ``frustrated synchronisation'' or ``Griffiths-like'' phase, without the explicit need for parameter fine-tuning to set parameter values close to the edge a bifurcation line or critical point. 

Let us finally mention that detailed analyses of the emerging short-time correlation matrices and their dynamical changes ---which can be very illuminating to investigate the segregation-integration balance--- are still missing. Further systematic analyses of the temporal structure of dynamical functional networks (following, e.g., the pioneering work of Deco's group \cite{Deco-discrete1,Deco-discrete2}) are in progress and will be reported elsewhere. Additionally, we also leave for future work a detailed analysis of how the network architecture affects the statistics of avalanches nearby the different transition points; owing to structural heterogeneity we expect the emergence of broad Griffiths regions with continuously varying avalanche exponents \cite{GPCN1,GPCN2,Moretti}, but studies employing much larger network sizes would be required to confirm or exclude this possibility in models with either homogeneous or heterogeneous frequency distributions.

\section{Conclusions and Discussion}

Understanding what are the chief structural of dynamical aspects that confers their extremely rich dynamical repertoire to actual brain networks is a challenge of outmost importance. For instance, as argued in the Introduction, such a large dynamical repertoire is a necessary condition to support a rich and flexible balance between segregation and integration.  In spite of very important advances ---both empirical and computational--- some aspects still remain obscure and/or controversial. In this brief paper, we have proposed a strategy to explore the problem of the emerging collective variability of functional and transient synchronisation: it consists on the consideration of simple structural networks and minimal dynamical models for the neural activity at single nodes (coupled together following the underlying network connectivity) with the goal of  scrutinising in a systematic way the emerging variability in synchronisation levels across time and network scales \new{and how it depends on key features of both the structure and  the dynamics.}

The results presented here constitute a first approximation to this ambitious long-term project;  much more extensive and systematic analyses along similar lines are still needed and will be presented elsewhere. Nevertheless, there are a number of features that become clear already from the present study. Among the few network architectures analysed here, those with a hierarchical-modular structure exhibiting also connector hubs, i.e. a core-periphery structure, are those supporting the largest possible range of dynamical variability.  Other structural aspects such as connectivity weights \cite{Bassett-weighted} and directionality \cite{Joglekar} are known to play a key role and need to be incorporated and studied in future extensions of this work. On the other hand, from the dynamical side, the main conclusion is that large levels of variability can be obtained in broad regions of parameter space, corresponding to Griffiths-like phases of ``frustrated synchronisation'', where transient and variable levels of synchrony are observed.  If one insists in fine-tuning the dynamics, the largest variability is obtained nearby hybrid-type transitions. 

We plan to extend the present work in a number of directions.  First of all, we intend to analyse structural matrices (including empirical ones) employing spectral graph theory, to understand their key features in a more systematic and consistent way. This will allow us to shed light on the structural modes that the topology supports (see, e.g., \cite{Atasoy1,Atasoy+Carhart}). The key idea is to determine at a graph-spectral level which are the dynamical regimes leading to the broader and more variable spectrum of excitations (see e.g. \cite{Zhou-jointly}).  We  also plan to analyse more detailed dynamical models \new{(beyond simple phase models)} such as the Landau-Ginzburg mesoscopic theory and Wilson-Cowan models \cite{LG,WC}. For example, it has been recently found that chaotic deterministic oscillators can generate a richer and more robust dynamical variability that stochastic system siting at the edge of a bifurcation \cite{Piccinini}. We plan to explore this possibility by analysing more neuro-physiologically motivated chaotic systems in the spirit of the Landau-Ginzburg approach \cite{LG, Cortes2012}. Other potentially important aspects, such as (i) time-delays in the node couplings, (ii) more realistic frequency distributions (as well as their correlation with node-connectivity and network location \cite{Gollo+Pang}, etc.) could also be considered. It is our hope that this type of modelling-oriented systematic approaches will help shedding light onto the rich collective dynamical regimes sustaining the functionality of actual brains, \new{bridging the gap between structure and dynamically emergent high-level functions}.

\section*{Methods}

\subparagraph{Synthetic network architectures} HM-R networks and HM-CP are built employing the algorithms proposed in \cite{Moretti} and \cite{Zamora}, respectively. Both models generate hierarchical-modular structures and are flexible enough to produce diverse networks depending on parameters such as the number of hierarchical levels, the total connectivity and the number of nodes in the basal modules. The HM-R case has two large-scale moduli (``hemispheres''), which on their term are composed by two sub-moduli, and so on for a total of $6$ hierarchical levels. The resulting $64$ basal moduli are ER networks with $N=25$ units and connectivity per node $k_0=12$ (see \cite{Moretti}). The next first hierarchical layer over the basal one has $k=5$, which is progressively reduced up to  a value $k=0.1$ for nodes connecting both hemispheres. The HM-CP network is constructed with identical basal moduli, the same number of layers and identical averaged connectivity, using the scale-free exponent $\gamma=2$ for all levels except the basal one \cite{Zamora}.

\subparagraph{Fano factor} The Fano Factor of a  time series $X(t)$ (either continuous of discrete in time) is defined as the ratio between the variance and the mean of the series, as $\text{FF} = \var(X)/\langle X \rangle$.  We computed the Fano factor over interspike times in the raster plot \cite{Dayan}, where a spike is defined each time the oscillator phase crosses $2\pi$. For each individual unit, the interspike interval is the time between two consecutive events in the raster plot, i.e., $\isi_j = t_{j+1} - t_{j}$ for $j=1,\ldots,n_\text{spikes}$. This is repeated for each unit, $\text{FF}_i$ for $i=1,\ldots,N$, computing then the histogram averaged over all oscillators.



\aucontribute{All authors conceived and designed the study, and they
  all drafted, read and approved the manuscript. VB and PV carried
  out the computational simulations and numerical analyses. MAM and RB supervised the research.}

\competing{The authors declare that they have no competing interests.}

\funding{MAM acknowledges the Spanish Ministry and Agencia Estatal de
  investigaci{\'o}n (AEI) through grant $FIS2017-84256-P$ and $PID2020-113681GB-100$ (European
  Regional Development Fund), as well as the Consejeria de
  Conocimiento, Investigaci{\'o}n Universidad, Junta de Andalucia and
  European Regional Development Fund, Ref. $A-FQM-175-UGR18$ and
  $P20-00173$ for financial support.  V.B. acknowledges support from the Sofja Kovalevskaja Award
from the Alexander von Humboldt Foundation, endowed by the Federal Ministry of Education and Research. 
  R.B. acknowledge funding from the INFN BIOPHYS project.
  We also thank Cariparma for their support through the TEACH IN PARMA project.}

\ack{We thank G. B. Morales and J. Pretel for extremely
  valuable comments and P. Moretti, J. Cortes, and S. di Santo for a long-term collaboration on the topics discussed here.}








\end{document}